\documentclass[prb,12pt]{revtex4}
\usepackage{amsmath}    % need for subequations
\usepackage{graphicx}   % need for figures
\usepackage{verbatim}   % useful for program listings
\usepackage{color}      % use if color is used in text
\usepackage{subfigure}  % use for side-by-side figures
\usepackage{hyperref}   % use for hypertext links, including those to external documents and URLs

\begin{document}

\title{Low-temperature anharmonicity of barium titanate: a path-integral molecular dynamics study}

\author{Gr\'egory Geneste}
\email[Corresponding author: ]{gregory.geneste@cea.fr}
\affiliation{CEA, DAM, DIF, F-91297 Arpajon, France}

\author{Hichem Dammak}
\affiliation{Laboratoire Structures, Propri\'et\'es et Mod\'elisation des Solides, CNRS UMR 8580, Ecole Centrale Paris, F-92295 Ch\^atenay-Malabry, France}
\affiliation{Laboratoire des Solides Irradi\'es, Ecole Polytechnique, CEA-DSM, CNRS, F-91128 Palaiseau, France}

\author{Marc Hayoun}
\affiliation{Laboratoire des Solides Irradi\'es, Ecole Polytechnique, CEA-DSM, CNRS, F-91128 Palaiseau, France}

\author{Mickael Thiercelin}
\affiliation{Laboratoire Structures, Propri\'et\'es et Mod\'elisation des Solides, CNRS UMR 8580, Ecole Centrale Paris, F-92295 Ch\^atenay-Malabry, France}
   \date{\today}

    \pacs{02.70.Ns, 63.20.Ry, 77.84.-s}

\begin{abstract}
We investigate the influence of quantum effects on the dielectric and piezoelectric properties of barium titanate in its (low-temperature) rhombohedral phase, and show the strongly anharmonic character of this system even at low temperature. For this purpose, we perform path-integral molecular-dynamics simulations under fixed pressure and fixed temperature, using an efficient Langevin thermostat-barostat, and an effective hamiltonian derived from first-principles calculations. The quantum fluctuations are shown to significantly enhance the static dielectric susceptibility ($\approx$ by a factor 2) and the piezoelectric constants, reflecting the strong anharmonicity of this ferroelectric system even at very low temperature. The slow temperature-evolution of the dielectric properties observed below $\approx$ 100~K is attributed 
(i) to zero-point energy contributions and 
(ii) to harmonic behavior if quantum effects are turned off.
\end{abstract}

\maketitle

\section{Introduction}

It is a common result of solid state physics that, below its Debye temperature, $\theta_{\text D}$, a solid  exhibits a behavior that significantly deviates from the predictions of classical mechanics, because of quantum fluctuations associated to atomic motions: in such conditions, one or several vibration modes of the system do not behave classically, {\it i.e.} the energy quantum $\hbar \omega$ separating their eigenstates is significantly higher than the thermal energy $k_B T$.
In harmonic, or mildly anharmonic systems, quantum effects can be theoretically treated in the framework of the harmonic or quasiharmonic approximation, using the standard quantization of atomic vibrations through the introduction of the normal coordinates, and many solids can be considered as harmonic below a certain temperature. Such harmonic systems usually exhibit weak dependence of their dielectric properties with temperature, and their dielectric permittivity is not sensitive to quantum effects\cite{epsilon}.

However, the influence of quantum effects is more complex in systems in which the microscopic degrees of freedom move inside a strongly anharmonic energy landscape. For example, in a number of crystals exhibiting a form of ferroic order related to atomic displacements (ferroelectric, ferroelastic, antiferrodistortive, ferrotorroidic, etc), most of the degrees of freedom typically evolve throughout a multiple-well energy surface. If the system is considered as classical (quantum effects neglected), there always exists a temperature below which such system exhibits harmonic behavior (it is finally trapped in one single well of the energy surface), but in a quantum system, if the quantum fluctuations on the displacements in the ground state extend beyond the harmonic region associated to each well, the standard treatment of quantum effects by means of the harmonic approximation becomes irrelevant. In such cases, {\it where quantum fluctuations of the ground state cause the system to probe the anharmonic part of the potential}, deviations from the harmonic approximation are expected down to zero~K.

Ferroelectric systems are typically characterized by complex multiple-well energy landscapes affecting the polar degrees of freedom, and the interplay between quantum fluctuations and polar modes at low temperature is strongly system-dependent. In the case of a deep multiple-well surface, the effects might be limited to weak quantum delocalization. But in the case of a shallow multiple-well surface, dramatic consequences can be observed, up to strong tunneling effect that can be able to fight against the order parameter, and even make it disappear. One of the most spectacular interaction between quantum fluctuations and polar degrees of freedom is indeed the suppression of the ferroelectric polarization due to quantum zero point motions in the so-called "quantum paraelectric" materials such as SrTiO$_3$\cite{muller,sto1} and KTaO$_3$,\cite{kto} two systems in which the existence of an underlying ferroelectric instability (thus associated to a shallow multiple-well energy surface) is experimentally infered from the very large and saturated values of the dielectric permittivity just above zero~K, and suggested from density-functional theory (DFT) calculations in the case of SrTiO$_3$\cite{sai}. Similar effects are suggested in another family of compounds, the "high-temperature" quantum paraelectrics, such as CaTiO$_3$\cite{kim,burton}, La$_{1/2}$Na$_{1/2}$TiO$_3$\cite{inaguma,lnto} or in a more general way RE$_{1/2}$Na$_{1/2}$TiO$_3$,\cite{LnNTO} in which RE features a rare-earth element. However, even in "standard" ferroelectric crystals such as BaTiO$_3$ (BTO), the quantum effects have been shown by path-integral Monte-Carlo technique to significantly decrease the phase transition temperatures by $\approx$ 30-50~K\cite{sto1} and to strongly modify the shape of the pressure-temperature phase diagram\cite{sto2} up to room temperature. Quantum effects can thus strongly influence the structural and dielectric properties of ferroelectric (FE) systems, not only at low temperature but also at room temperature and beyond.

Anharmonicities in BTO are impacting the physics even at low temperature, as shown by its lowest phase transition (rhombohedral-orthorhombic -- expt: 183~K\cite{tcexpt}, $\approx$ 190~K with the hamiltonian used in the present work\cite{zhong95}). In a simple picture, each phase transition in BTO corresponds, upon heating, to the temperature at which the local modes (dipoles) get out of the potential well(s) in which they were confined, and come to visit new potential energy minima, giving rise to a new value and direction of the macroscopic polarization. In other words, phase transitions are the points at which anharmonicities in the potential energy surface strongly manifest and impact the physics. This simple microscopic picture of BTO has been theorized more than 40 years ago through the well-known "eight-site" order-disorder model of Comes {\it et al.}\cite{comes}. In this model, the local dipoles evolve among eight off-center satellite sites located along the $<$111$>$-type directions. In the paraelectric phase, all the sites are visited with equal probability (resulting in a zero polarization), while in the FE phases, only a subset of these sites (the same at all cells due to strong intersite correlations) is visited, giving rise to a non-zero polarization. Modern calculations using an effective hamiltonian fitted on first-principles calculations\cite{zhong95} show that, in the paraelectric phase, the local density of probability is in fact quasi-isotropic (with slight maxima along the $<$111$>$ directions) since the dipoles also spend a significant part of their time between these off-center sites\cite{correl,correl2}. At any rate, this simple model confirms the strongly anharmonic character of this system, at least above its first phase transition and thus, {\it the complete impossibility to describe its phase transitions in the harmonic approximation}. 
Let us mention that the signature of anharmonicities is seen also below the first phase transition: in the rhombohedral phase, order-disorder mechanims associated with local reversal of the dipoles have been observed, both experimentally\cite{volkel2007} and by calculations,\cite{correl,correl2} suggesting anharmonic behavior also in the ground state phase of BTO.

On the other hand, the Debye temperature of BTO is commonly placed about 150~K above room temperature ($\theta_{\text D} \approx$ 480~K\cite{Meng2010}), showing that the dipole dynamics should be impacted by quantum effects up to such high temperature. 
Therefore, since the structural phase transitions occur well below $\theta_{\text D}$, subtle interplay between quantum fluctuations and anharmonicities are expected in BTO. Let us also mention the fact that the so-called "temperature rescaling" method\cite{rescale} does not apply to BTO since the effective temperature calculated from the phonon density of states obtained at $T=0$~K in the rhombohedral phase is about 250~K, which falls above the two first phase transitions of BTO, namely in the tetragonal phase.

In the present work, we investigate the anharmonicities of barium titanate at low temperature (in its rhombohedral phase), that manifest through the quantum fluctuations of its ground state. We describe the influence of quantum effects on the order parameter, on the dielectric permittivity and on the piezoelectric constants of BTO, by using path-integral molecular-dynamics simulations.

\section{Computational details and theoretical background}
\subsection{Path-integral formalism}

The quantum effects related to atomic motions are accounted for by using the path integral (PI) formalism.
In this formulation of quantum statistical mechanics, the canonical partition function $Z$ is written as a discretized imaginary time path integral. 
For a quantum system containing $N$ (discernable) particles of mass $m$, $Z$ can be expressed according to:

\begin{equation}
\label{eq0}
Z = \lim_{P \rightarrow \infty} 
 (\frac{2 \pi m P k_B T}{h^2})^{3NP/2}  \times 
\int_{\vec U^{(1)}} ... \int_{\vec U^{(P)}} e^{-\beta V_{eff}(\{ \vec U^{(1)} ... \vec U^{(P)} \})} d \vec U^{(1)} ... d \vec U^{(P)}.
\end{equation}

This integral involves $P$ (Trotter number) replicas of the system labeled by the integer $s$, each replica consisting of a set of $N$ positions $\vec U^{(s)} = (\vec u_1^{(s)}  ...  \vec u_N^{(s)})$. These replicas characterize the discretization of the PI in imaginary time (imaginary time slices).
$\beta$ is the statistical temperature, $\beta = \frac{1}{k_B T}$. The effective potential $V_{eff} (\{ \vec U^{(1)} ... \vec U^{(P)} \})$ couples the positions of the $P$ slices through:

\begin{equation}
V_{eff}(\{ \vec U^{(1)} ... \vec U^{(P)} \}) = \sum_{s=1}^P [\sum_{i=1}^N \frac{1}{2} k (\vec u_i^{(s)} - \vec u_i^{(s+1)})^2 + \frac{1}{P} \Phi(\{ \vec u_i^{(s)} \})].
\end{equation}

The harmonic term of this effective potential involves a spring constant $k = \frac{m P k_B^2 T^2}{\hbar^2}$. 
$\Phi$ is the physical potential energy computed inside each imaginary time slice $s$. 
Each particle $(i,s)$ is thus interacting by harmonic forces with the particles $(i, s+1)$ and $(i,s-1)$, forming a ring that closes onto itself ($\vec u_i^{(P+1)} = \vec u_i^{(1)}$).

In the limit of infinite Trotter number, Eq.~(\ref{eq0}) tends to a functional integral (imaginary-time path integral):

\begin{equation}
Z= \oint D \vec U e^{- {\frac{1}{\hbar}} \int_{0}^{\beta \hbar} [ T(\frac{d \vec U}{d \tau}) + \Phi(\vec U(\tau)) ] d \tau },
\end{equation}

the integral being over all paths [$\tau \in [0; \beta \hbar] \rightarrow  \vec U(\tau) \in R^{3N}$] with the cyclic condition [$\vec U(0) =  \vec U(\beta \hbar)$], $T$ and $\Phi$ in the Euclidean action being the kinetic and potential energies, that respectively depend on the momenta $\frac{d \vec U}{d \tau}$ and positions $\vec U(\tau)$.

The discretized expansion of Eq.~\ref{eq0} is at the root of a formal analogy\cite{isomorphism} ("classical isomorphism"\cite{ceperley}) between any quantum system and an equivalent classical system made of $P$ images of the initial set of $N$ particles, because the multidimensional integral of Eq.~\ref{eq0} can be viewed as the canonical partition function of this equivalent classical system.
Each quantum particle is associated to a ring polymer of $P$ classical particles, these classical particles interacting with each other through the "true" physical forces (divided by $P$) inside each slice, and through harmonic forces (acting between each particle of slice $(s)$ and the corresponding particles of slice $(s-1)$ and $(s+1)$). Each set of harmonic interactions is assumed to close onto itself, forming a closed ring ($P+1 \rightarrow 1$). In the limit where the Trotter number $P \rightarrow \infty$, this equivalent classical system has exactly the same partition function as that of the quantum system under study. The extension to the isothermal-isobaric ensemble (NPT) is straightforward\cite{loubeyre}.

As a consequence, classical simulation techniques such as Monte Carlo (MC) or molecular dynamics (MD) can be applied to the classical equivalent to estimate numerically the thermodynamic properties of the quantum system. The corresponding methods are respectively called path-integral Monte Carlo and path-integral molecular dynamics (PIMD). The estimated properties should be converged with the Trotter number, and if this condition is fullfilled, it is possible to compute thermodynamic quantities that exactly include all the quantum {\it dispersion} effects. 
However, the path-integral expansion of Eq.~\ref{eq0} assumes distinguishible particles. The physical properties computed by the present PI formalism thus do not include exchange between particles, which are assumed to obey Boltzmann statistics\cite{marx}.

Practical application of PIMD raises technical problems, related to the fact that for high Trotter number, the forces acting inside the classical equivalent are mainly harmonic. In such harmonic system, obtaining ergodic trajectories by using the standard algorithms of MD such as Nos\'e-Hoover thermostat is difficult\cite{chains}. In order to recover ergodicity, two different approaches can be employed:

(i) a deterministic approach using efficient schemes based on {\it thermostat chains}\cite{chains,marx,tuckerman};

(ii) a stochastic approach based on the use of the Langevin thermostat\cite{lan1,lan2,lan3,lan4}.

In the present case, we have found very efficient to use the Langevin dynamics, which is extremely powerful to produce an ergodic exploration of phase space by introducing at each time step a random force that mimics the "noise" observed in the motion of a brownian particle. We have successfully tested the scheme on simple systems (1D and 3D harmonic oscillator, double well potential, quartic potential) and found an excellent agreement between PIMD and the exact result (obtained by analytic formulae or by a numerical solution of the Schr\"odinger equation).

%###################
Another practical difficulty of PIMD is the existence of modes evolving on very different time scales\cite{tuckerman}. To circumvent this difficulty, a specific coordinate transformation ("normal mode" or "staging") that diagonalizes the harmonic parts of the PIMD forces could be employed. An appropriate choice of fictitious masses gives the same time scale for the dynamics of each degree of freedom. Instead of using such a coordinate transformation, we performed long enough trajectories (see Sec.~\ref{md}).
%###################

\subsection{Hamiltonian}
We use the effective hamiltonian of Zhong {\it et al.}\cite{zhong95}, which is derived from first-principles density-functional calculations and has been shown to provide an excellent description of the thermodynamics of BTO, especially its complex sequence of phase transitions: rhombohedral(R) - orthorhombic(O) - tetragonal(T) - cubic(C) -- and the (first) order of its phase transitions. In particular, although the Curie temperature $T_c$ is predicted at $\approx$ 300~K, {\it i.e.} about 100 K too low with this hamiltonian, it provides a good value for the R-O phase transition temperature ($\approx$ 190~K without inclusion of quantum effects\cite{zhong95}, and  $\approx$ 160~K after inclusion of quantum effects, to be compared with the experimental value of 183~K\cite{tcexpt}). The degrees of freedom of this hamiltonian are the local modes $\{ \vec u_i \}$, the mechanical displacement modes $\{ \vec v_i \}$ and the (homogeneous) strain tensor $\{ \eta_l \}$. $\vec u_i$ is, roughly speaking, the local polar displacement inside cell $i$, related to the local dipolar moment $\vec d_i$ through an effective charge $Z^{*}$: $\vec d_i = Z^{*} \vec u_i$. For simplicity, the mechanical displacement modes -- that allow appearance of an inhomogeneous component of the total strain and have been shown to be of very weak influence in this material\cite{zhong95} -- are not accounted for in the present study. The hamiltonian  $\Phi(\{ \vec u_i \}, \{ \eta_l \})$ is thus a function of the local modes $\{  \vec u_i \}$, and of the (homogeneous) strain tensor components $\{ \eta_l \}$. It consists of a (local) "onsite" part, a long-range dipole-dipole interaction term, a term describing short-range interactions between neighboring local modes (up to 3$^{rd}$ neighbor), a (local) term that couples the local mode to the strain and an elastic energy\cite{zhong95}. In the presence of an external electric field $\vec E$, a term $- \sum_{i} Z^{*} \vec u_i . \vec E$ is added.

In this work, we use in the discussion the mean local mode $<$$\vec u$$>$ as order parameter (this is a displacement). An unambiguous relationship with the spontaneous polarization $\vec P$ can be made by $\vec P = Z^{*} \frac{ < \vec u >}{\Omega}$, where $\Omega$ is the unit cell volume and $Z^{*}$ the effective charge associated with the local mode (9.956 $e$). The components of $<$$\vec u$$>$ are expressed in $a_0$ units, where $a_0$ is the theoretical lattice constant of BTO\cite{zhong95} computed in the framework of the local-density approximation to DFT: $a_0$ = 7.46 Bohrs.
For each temperature, the macroscopic order parameter $<$$\vec u$$>$ and the homogeneous strain are obtained by averaging over unit cells, (real) time steps, and imaginary time slices, allowing to determine the symmetry of the phase (R, O, T or C).

\subsection{Langevin barostat}
BTO is simulated under fixed (hydrostatic) pressure and fixed temperature conditions. Since Langevin dynamics is very efficient to recover ergodicity within the PI formalism in the canonical ensemble, we wish to use this method, not only under fixed temperature but also under fixed pressure. The extension of the Langevin method to the isothermal-isobaric ensemble has precisely been achieved by Quigley and Probert\cite{qp1,qp2}, giving rise to an algorithm in which random and friction forces are applied, not only on the atomic coordinates, but also on the supercell vectors. We have thus implemented this "Langevin barostat" within the PI formalism. In what follows, second-rank tensors are written in bold. The equations of motion on local mode $i$ of slice $(s)$ (with mass $m$) using the Langevin barostat are:

\begin{equation}
\frac{ d \vec p_i^{(s)} } { dt} = \vec f_i^{(s)} - \gamma  \vec p_i^{(s)}  + \vec L_i^{(s)} - \frac{\bf p_G}{W_g} \vec p_i^{(s)} - \frac{1}{N_f} . \frac{Tr({\bf p_G})}{W_g} \vec p_i^{(s)},
\end{equation}

with $\vec f_i^{(s)} = - \frac{1}{P} \vec \nabla_{\vec u_i^{(s)}} \Phi(\vec u_1^{(s)},...,\vec u_N^{(s)}) - k(T,P)(2 \vec u_i^{(s)} - \vec u_i^{(s+1)} - \vec u_i^{(s-1)})$ the PIMD force that includes the quantum kinetic energy contribution, which takes the form of an harmonic force with spring constant $k(T,P)= m P k_B^2 T^2 / \hbar^2$. The term $- \gamma \vec p_i^{(s)} $ corresponds to the friction force of the Langevin thermostat and $\vec L_i^{(s)}$ is the so-called Langevin force, which is randomly drawn at each time step in a gaussian of variance $\sqrt{\frac{2 \gamma m k_B T}{\delta t}}$, $\delta t$ being the time step.
The momentum $\vec p_i^{(s)}$ is related to the position $\vec u_i^{(s)}$ by

\begin{equation}
\frac{d \vec u_i^{(s)}}{dt} = \frac{\vec p_i^{(s)}}{m} + \frac{\bf p_G}{W_g} \vec u_i^{(s)},
\end{equation}

while the matrix of the supercell vectors ${\bf h}$ and its conjugate momentum ${\bf p_G}$ evolve according to

\begin{equation}
\frac{d{\bf h}}{dt} = \frac{\bf p_G h}{W_g},
\end{equation}

and

\begin{equation}
\label{eq1}
\frac{d {\bf p_G}}{dt} = V(t)({\bf X} - P_{ext} {\bf Id}) + \frac{1}{N_f} \sum_{i,s} \frac{ {\vec p_i^{(s)^2}} } {m}  {\bf Id} - \gamma_G  {\bf p_G} + {\bf L_G},
\end{equation}

in which $V(t)$ is the supercell volume (that evolves with time), $W_g$ is the "mass" associated to the barostat, $N_f$ is the number of degrees of freedom, $P_{ext}$ is the external pressure, ${\bf Id}$ is the identity tensor and ${\bf X}$ is the internal pressure tensor\cite{qp1,martyna1999}.
In the right member of Eq.~\ref{eq1}, one recognizes a friction force on the supercell $- \gamma_G  {\bf p_G}$ ($\gamma_G$ is a friction coefficient for the barostat) and a random force ${\bf L_G}$, a 3 $\times$ 3 matrix whose components are randomly drawn at each time step in a gaussian with variance  $\sqrt{\frac{2 \gamma_G W_g k_B T}{\delta t}}$. This random force on the barostat is symmetrized at each time step to avoid global rotation of the supercell during the simulation. Let us precise that the same supercell ${\bf h}$ is common to all the imaginary time slices (there is no replication of the supercell).

The Langevin algorithm is very sensitive to the quality of the random number generator. In this study, we have adopted the routines of R. Chandler and P. Northrup\cite{rand1,rand2}. Finally, the algorithm is very stable and thermalization is fully achieved within $\approx$ 20000 MD time steps.

\subsection{Molecular dynamics}
\label{md}
The MD simulations are performed using a 12$\times$12$\times$12 supercell with periodic boundary conditions. The time step is $\delta t$=1.0$\times$10$^{-15}$ s. The mass associated to the local mode (not important in the classical case for the computation of ensemble averages, but crucial in the quantum case) is 39.0 atomic mass units, determined from the force constant matrix eigenvector defining the local mode. This value is the same as that used in Ref.~\onlinecite{nishimatsu}. The external pressure is fixed at -4.8 GPa, as in Ref.~\onlinecite{zhong95}, a negative value that compensates the underestimation of the lattice constant within the local-density approximation to DFT.
The Langevin-PIMD equations of motion are integrated within the Verlet algorithm: at each step, the new positions at time $t + \delta t$ are computed from the positions at $t$ and $t - \delta t$, and also from the velocities at $t$, as required by the equations of motion of the Langevin thermostat-barostat. These velocities are calculated self-consistently (a short internal loop is performed at each step of the Verlet algorithm) starting from an estimation of the velocity taken from Ref.~\onlinecite{ferrario}.

Convergence with the number of imaginary time slices has been carefully studied. Our tests show that various properties (polarization, strain and phase transition temperatures) of the system from $T=120$ to 300~K do not exhibit significant change from $P=8$ to $P=16$.
Thus the computation of the dielectric and piezoelectric tensors is achieved at low temperature (from $T=30 K$ to $T=137 K$) by maintaining $P\times T = 120\times 16 =  1920$, leading to the use of Trotter numbers as large as $P=64$ at the lowest temperature studied ($T=30$ K). Note that classical mechanics is recovered by setting the Trotter number $P$ to 1.

Equilibrium trajectories of typically 3$\times$10$^{5}$ (low T) up to 5$\times$10$^{5}$ (high T) steps are generated at each temperature after an equilibration time that can be long at low temperature, due to a very slow dynamics. Thus, at low temperature ($T\leq60~K$), an electric field along [111] is applied during the equilibration procedure to help the system reach faster its equilibrium state. The friction coefficient of the Langevin thermostat is $\gamma = $ 0.5 THz.

Dielectric and piezoelectric tensors have been computed using a finite differences method, by directly applying a finite electric field along [001]. Equilibrated trajectories of  PIMD steps are generated under static electric field and fixed pressure (-4.8 GPa). Different values of the electric field, in the range $[-5 , + 5 ]\times 10^{6}$ V/m, are chosen for each temperature. The field is weak enough to induce a linear dielectric response, in both the polarization and  the strain.

%################################
In order to ensure the sampling accuracy, we compare our PIMD averages with the one obtained by PIMD under the staging mode transformation. 
Since the difficulty to obtain such a good sampling could occur with high Trotter number, let us consider the most unfavourable case ($P=64$ at $T=30 K$).
A long $NPT$ trajectory has thus been followed by a $\approx$ 100000 step-$NVT$ trajectory performed by fixing the strain tensor components to the mean values deduced from the $NPT$ trajectory. This second trajectory was computed by using both the primitive coordinates, and the staging coordinates.
This allows to demonstrate that long trajectories, even using primitive coordinates, provide very well converged results, as shown in Tab.~\ref{staging}.
It is important to mention that the primitive averages converge to the staging ones after $\approx$ 20000-30000 times steps, whereas our $NPT$ trajectories contain more than 300000 time steps.
%#################################

\begin{table}[htdp]
\caption{Averaged local modes and stress at $T=30$ K and for $P=64$ obtained over
(i) a $NPT$ trajectory with primitive coordinates,
(ii) a $NVT$ trajectory with primitive coordinates,
(iii) a $NVT$ trajectory with staging coordinates.
The strain fixed for the $NVT$ cases is the one averaged along the $NPT$ trajectory.}
\begin{center}
\begin{tabular}{ccccc}
\hline
\hline
      &      &   $E_x=E_y=E_z=0$  &              &            \\
\hline
 Ensemble      &          &  $NPT$     &     $NVT$     &    $NVT$     \\
 Coordinates   &          &  prim.   &   prim.     &   staging  \\
 Local mode ($a_0$) &   $u_z$  &  0.021612& 0.021592    &   0.021604  \\
\hline
Stress (GPa)  &   $\sigma_1$  &  4.8 & 4.800383    &   4.802153 \\
              &   $\sigma_4$  &  0.0 & 0.000117    &  -0.000038   \\
\hline
               &          & $E_x=E_y=0$;  $E_z=6.4\times 10^6 $ V/m  &     &      \\
\hline
 Ensemble      &          &  $NPT$     &     $NVT$     &    $NVT$     \\
 Coordinates   &          &  prim.   &   prim.     &   staging  \\
Local mode ($a_0$)  &   $u_x$  &  0.021110& 0.021127    &   0.021125  \\
               &   $u_z$  &  0.022722& 0.022729    &   0.022731 \\
\hline
 Stress (GPa)  &   $\sigma_1$  &  4.8 & 4.798115    &   4.800640 \\
               &   $\sigma_4$  &  0.0 & -0.000025    &  -0.000031   \\
               &   $\sigma_6$  &  0.0 & -0.000059    &   0.000065  \\
\hline
\end{tabular}
\end{center}
\label{staging}
\end{table}

\section{Spontaneous polarization}
As a first step, we investigate the phase sequence of BTO by both classical and quantum simulations in order 
(i) to evaluate the importance of the quantum contributions as a function of temperature, and 
(ii) to examine whether the low-temperature spontaneous polarization is impacted by quantum fluctuations, providing a first idea of low-temperature anharmonicities. 

\begin{figure*}[htbp]
    {\par\centering
 {\scalebox{0.5}{\includegraphics{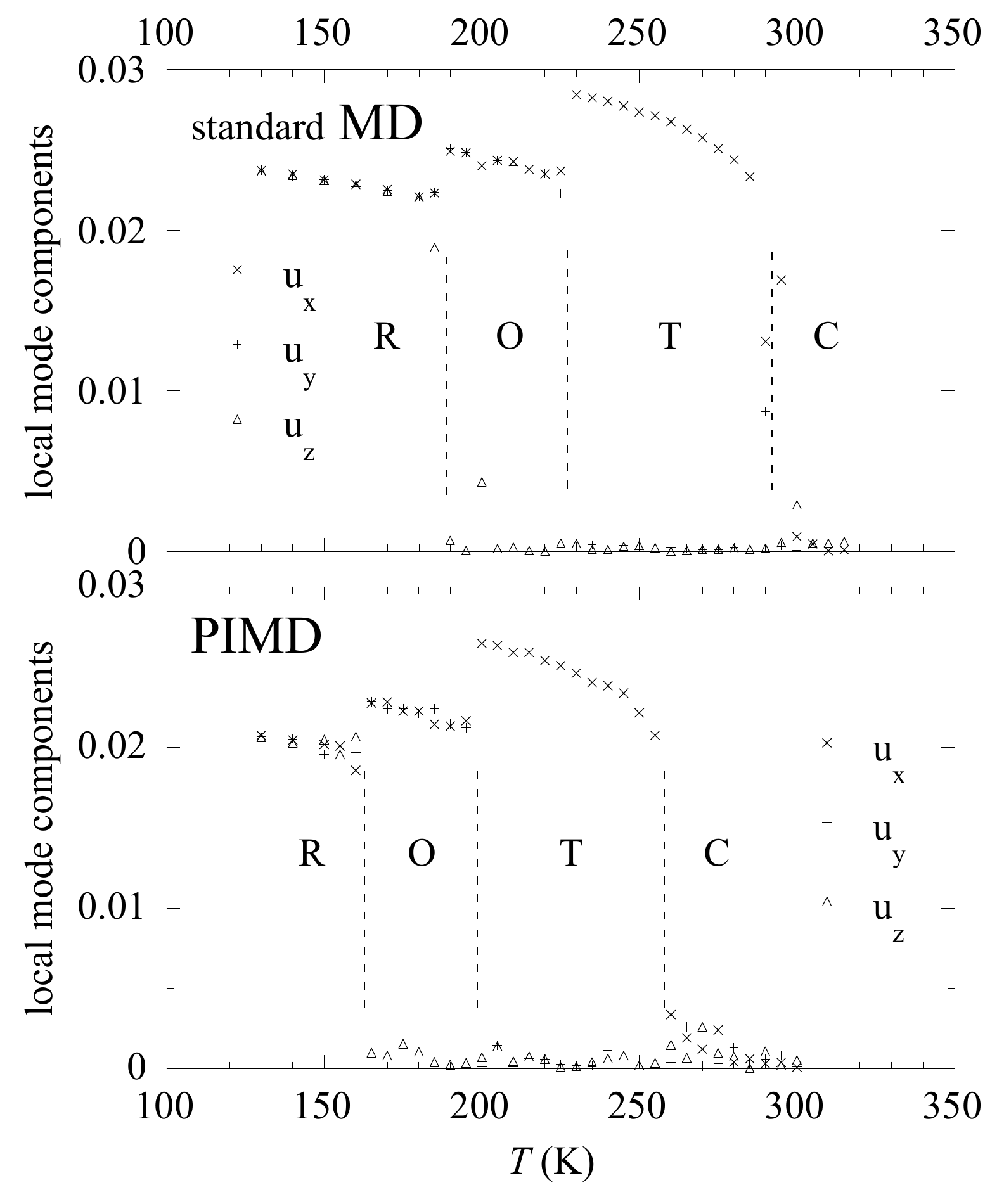}}}
    \par}
     \caption{{\small Temperature evolution of the mean local mode components (proportional to the polarization) as obtained by standard MD and PIMD.}}
    \label{Fig-pola-MD-PIMD}
\end{figure*}

In the classical case ($P=1$), the temperature evolution of polarization is displayed on Fig.~\ref{Fig-pola-MD-PIMD}. 
We find the expected sequence of phase transitions: R - O - T - C, as experimentally observed \cite{ROTCexp} and in excellent agreement with the classical Monte carlo calculations of Zhong {\it et. al.}\cite{zhong95} using the same hamiltonian. In particular, the transition temperatures are very close to those obtained by these classical Monte Carlo simulations\cite{zhong95}: the Curie temperature is found at $\approx$ 295~K, while the R-O and O-T transition points are obtained at $\approx$ 190~K and $\approx$ 230~K, respectively. These preliminary calculations illustrate the accuracy of the Langevin barostat to sample the $(NPT)$ ensemble. Note that these transition points are subject to uncertainties of $\approx$ 10-15~K related to hysteretic phenomena common to the simulation of first-order phase transitions. A more accurate determination of the phase transition temperatures would require to achieve a finite-size scaling analysis\cite{challa1986} which is beyond the scope of the present study.

In the quantum case, the transition temperatures (see Fig.~\ref{Fig-pola-MD-PIMD}) are lowered compared to the classical case by $\approx$ 10-15\%, and agree very well to the ones obtain by path-integral Monte Carlo by Zhong and Vanderbilt\cite{sto1}. This behavior is expected since the quantum fluctuations destabilize the ferroelectric order.
We also observe an important decrease in the spontaneous polarization in the three ferroelectric phases (Fig.~\ref{Fig-pola-MD-PIMD}). At $T=30$~K, we find in the quantum case $\left< u_x \right> = \left<u_y\right> = \left<u_z\right> =$ 0.0216~a$_0$  (0.221~C/m$^2$), whereas in the classical case, we have 0.0257 a$_0$ (0.263~C/m$^2$). This difference reflects the fact that {\it the mean polar displacement in the ground state is not at the minimum of the potential energy}. This is due to the quantum fluctuations extending to the asymmetric (and thus anharmonic) region of the potential energy surface of the system. Indeed, an harmonic system would have its ground state symmetric with respect to the minimum energy point. Despite the multidimensional character of this energy surface, it is possible to have an idea of its anharmonicities by plotting the energy as a function of polarization $\vec P$, with all the local modes fixed to the same value in all the unit cells of the system: $\vec u_1 = \vec u_2 = ... = \vec u_N = \vec u$. This simplified static energy landscape is displayed in Fig.~\ref{fig4} along three different directions and help understanding the effect of quantum fluctuations: since the energy surface is below (resp. above) the harmonic approximation along the [111] direction when the polarization is decreased (resp. increased) with respect to its value in the energy minimum, the quantum fluctuations decrease the value of the spontaneous polarization, as found in the calculations.

This significant difference between the classical and quantum systems is a signature of the importance of anharmonicities in the rhombohedral phase of BTO. In other words, quantum fluctuations lower the spontaneous polarization at low temperature by 15-20\%. This manifestation of quantum mechanics takes place through the strong anharmonicity of the potential energy surface of BTO.

\section{Dielectric permittivity}
We now focus on the rhombohedral phase of BTO below 140~K and compute the dielectric and piezoelectric tensors in the presence of quantum fluctuations. This calculation is achieved, as explained in the computational part, by applying a finite external electric field $\vec E$ along the [001] direction. Fig.\ref{dielectric} shows the temperature evolution of the transverse and longitudinal components of the static dielectric tensor, computed classically ($P$=1) and quantum-mechanically ($P \times T$ = 1920). These components are systematically given, and discussed, in the rhombohedral reference system (see Appendix).

\begin{figure}[htbp]
    {\par\centering
     {\scalebox{0.5}{\includegraphics{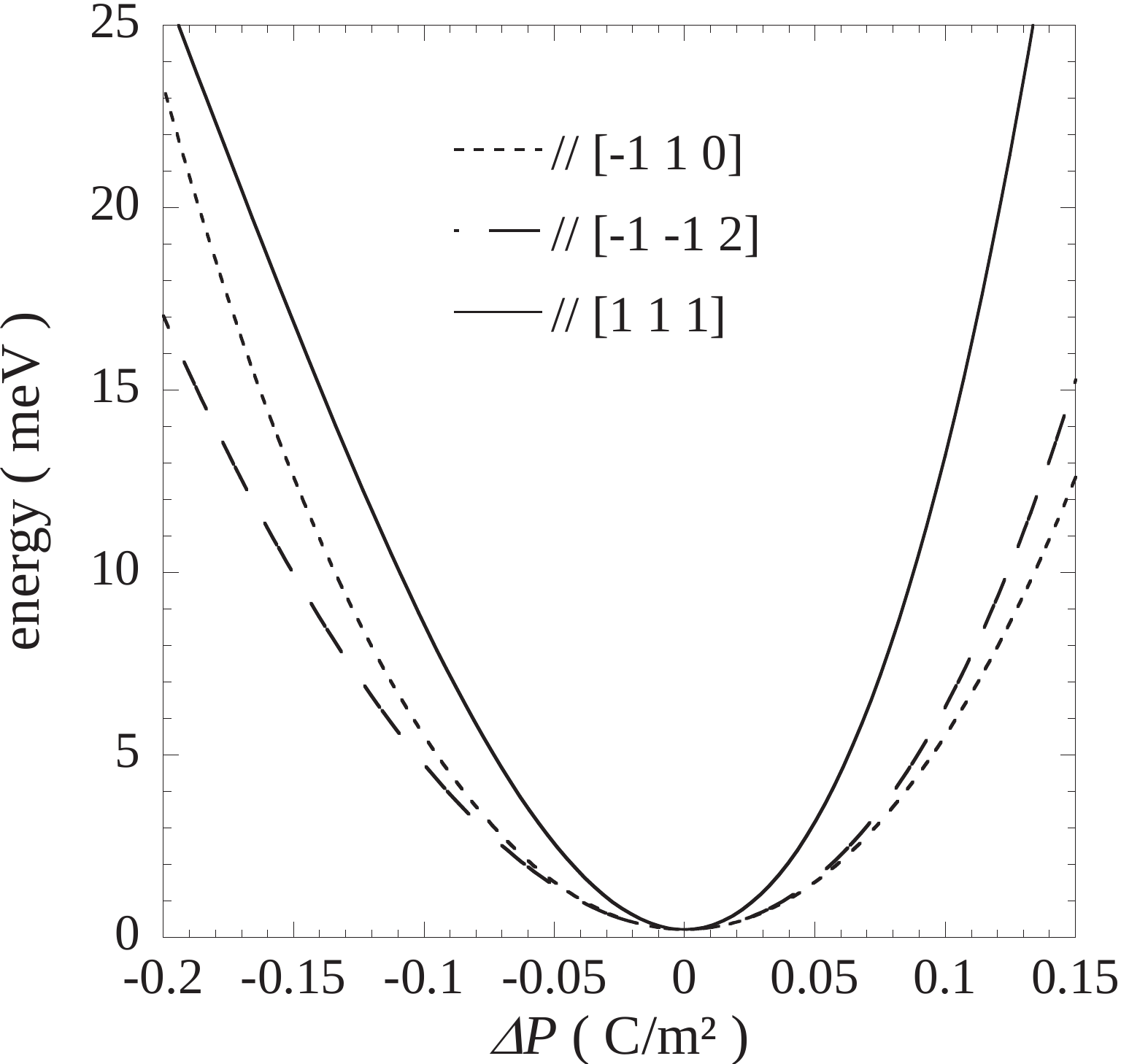}}}
    \par}
     \caption{Energy landscape of BTO in its rhombohedral phase as a function of the polarization variation around the energy minimum, along three directions.}
    \label{fig4}
\end{figure}

\begin{figure}[htbp]
    {\par\centering
     {\scalebox{0.5}{\includegraphics{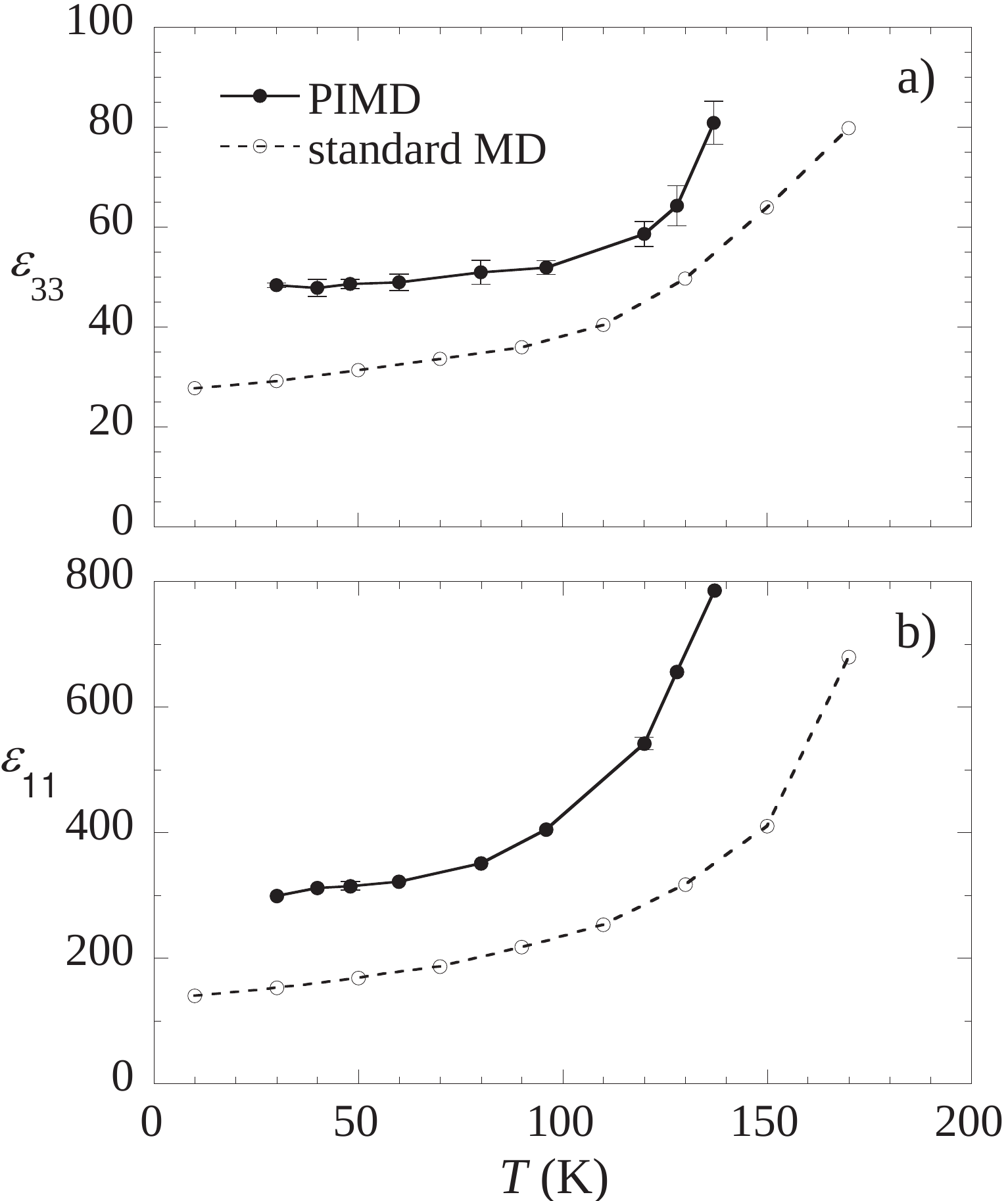}}}
    \par}
     \caption{{\small Temperature evolution of the longitudinal and transverse components of the dielectric tensor, computed using either standard MD or PIMD. The electronic contribution $\epsilon^{\infty}$ ($= 5.24$ in cubic BTO\cite{zhong95}) is not included.}}
    \label{dielectric}
\end{figure}

First, we note that, independently from the inclusion of quantum fluctuations, the transverse component, $\epsilon_{11}$, is much larger than the longitudinal component, $\epsilon_{33}$. This behavior is expected and in good agreement with the density functional perturbation theory (DFPT) calculations of Wu, Vanderbilt and Hamann\cite{wu}, who obtained $\epsilon_{11} = 265$ and $\epsilon_{33} = 50$ at $T=0$~K. The ratio between these two components can be explained through the curvature of the potential energy surface around the rhombohedral minimum. Fig.~\ref{fig4} shows how the energy landscape varies as a function of the polarization around this minimum. The $[\bar{1}10]$ and $[\bar{1}\bar{1}2]$ directions give insight into $\epsilon_{11}$, the [111] direction into $\epsilon_{33}$. The much sharper increase observed along the longitudinal [111] direction (Fig.~\ref{fig4}) shows that $\epsilon_{33}$ is smaller (smaller polarization fluctuations) than $\epsilon_{11}$ along the transverse ones.

We now examine the influence of quantum fluctuations.
In both the classical and quantum-mechanical cases, our calculations show that the dielectric constants slowly evolve with temperature for temperatures below $\approx 100$~K  (Fig.\ref{dielectric}). This common feature is explained differently according to the case. In the classical approach, the system, as explained previously, eventually becomes harmonic at sufficiently low temperature, leading to the slow temperature-evolution in the dielectric response. In the quantum case, the system rapidly reaches its ground state upon cooling, generating a freezing (saturation) of all the physical quantities. The slow evolution of the dielectric response at low temperature is therefore attributed to the quantum zero-point effects. For higher temperatures ($>100$~K) the dielectric constants increase and become very large as approaching from below the R-O phase transition.
We systematically find that the inclusion of quantum fluctuations enhances the dielectric response, approximately by a factor 2 for both $\epsilon_{11}$ and $\epsilon_{33}$. Such differences between the quantum and classical descriptions are the signatures of a strong anharmonicity in the potential energy landscape of the rhombohedral phase of barium titanate, since in an harmonic system, the static dielectric tensor would not depend on the inclusion of quantum effects and would be independent on the temperature\cite{epsilon}.

\begin{figure}[htbp]
    {\par\centering
     {\scalebox{0.5}{\includegraphics{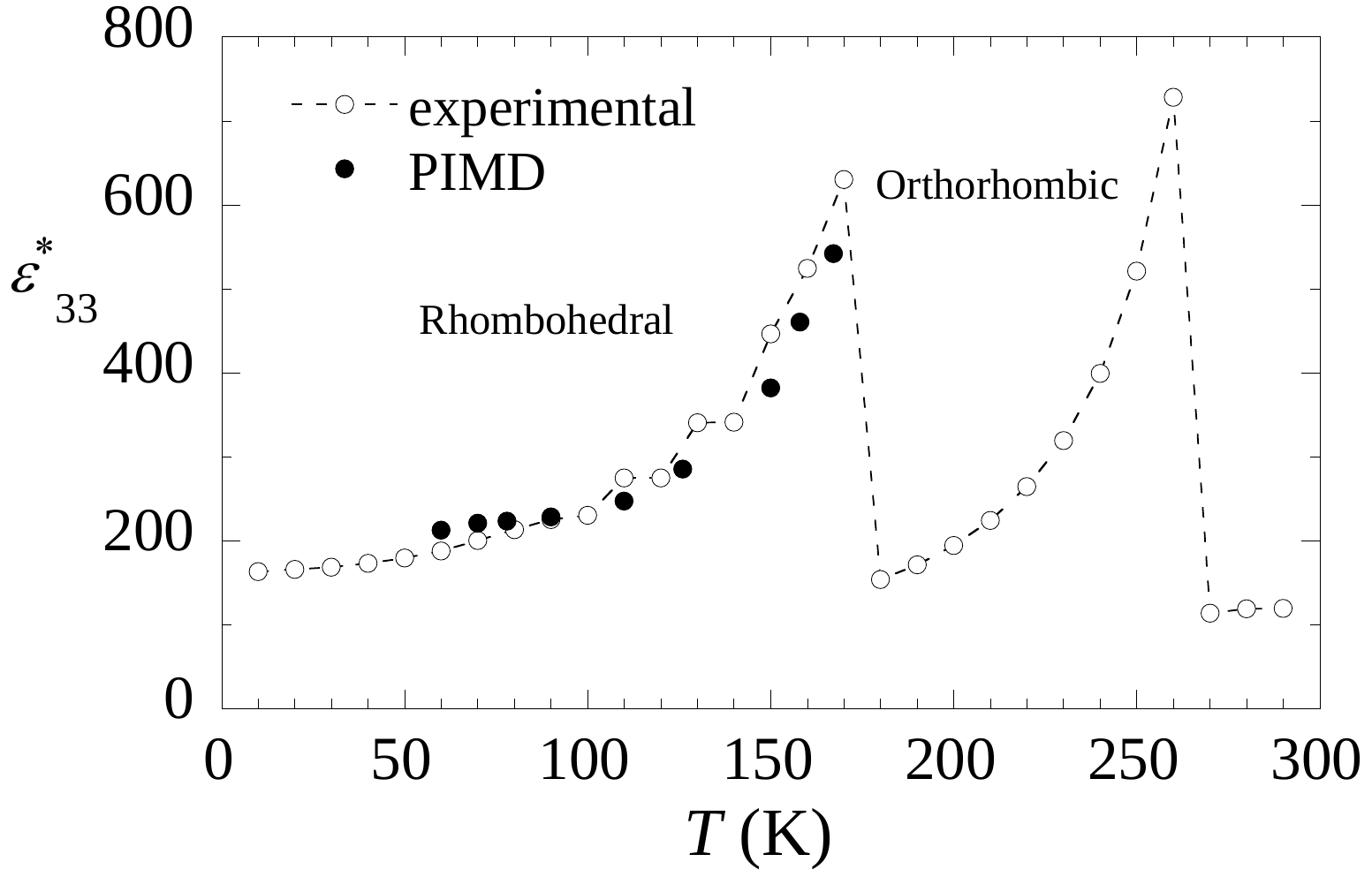}}}
    \par}
     \caption{{\small Temperature evolution of the longitudinal dielectric constant $\epsilon_{33}^*$ along $[001]$.}}
    \label{dielexp}
\end{figure}

Finally, we have also performed measurements of the dielectric constant, $\epsilon_{33}^*$, on a $[001]$ oriented single crystal in the 10--300~K temperature range by using an impedance analyzer. These results are given in Fig.~\ref{dielexp} and show two anomalies at $170\pm 5$ and $260\pm 5$~K corresponding to the R--O and O--T ferroelectric phase transitions. The PIMD $\epsilon_{33}^*$ (see Appendix) are shifted in temperature, to take into account the difference between our calculated and our experimental R-O transition temperatures. The agreement with the experiment is good, even though the PIMD values are overestimated at low temperatures and underestimated at high temperatures in the R phase. Hence, the effective hamiltonian provides a satisfactory behavior.

\section{Piezoelectric coefficients}
\begin{figure}[htbp]
    {\par\centering
     {\scalebox{0.5}{\includegraphics{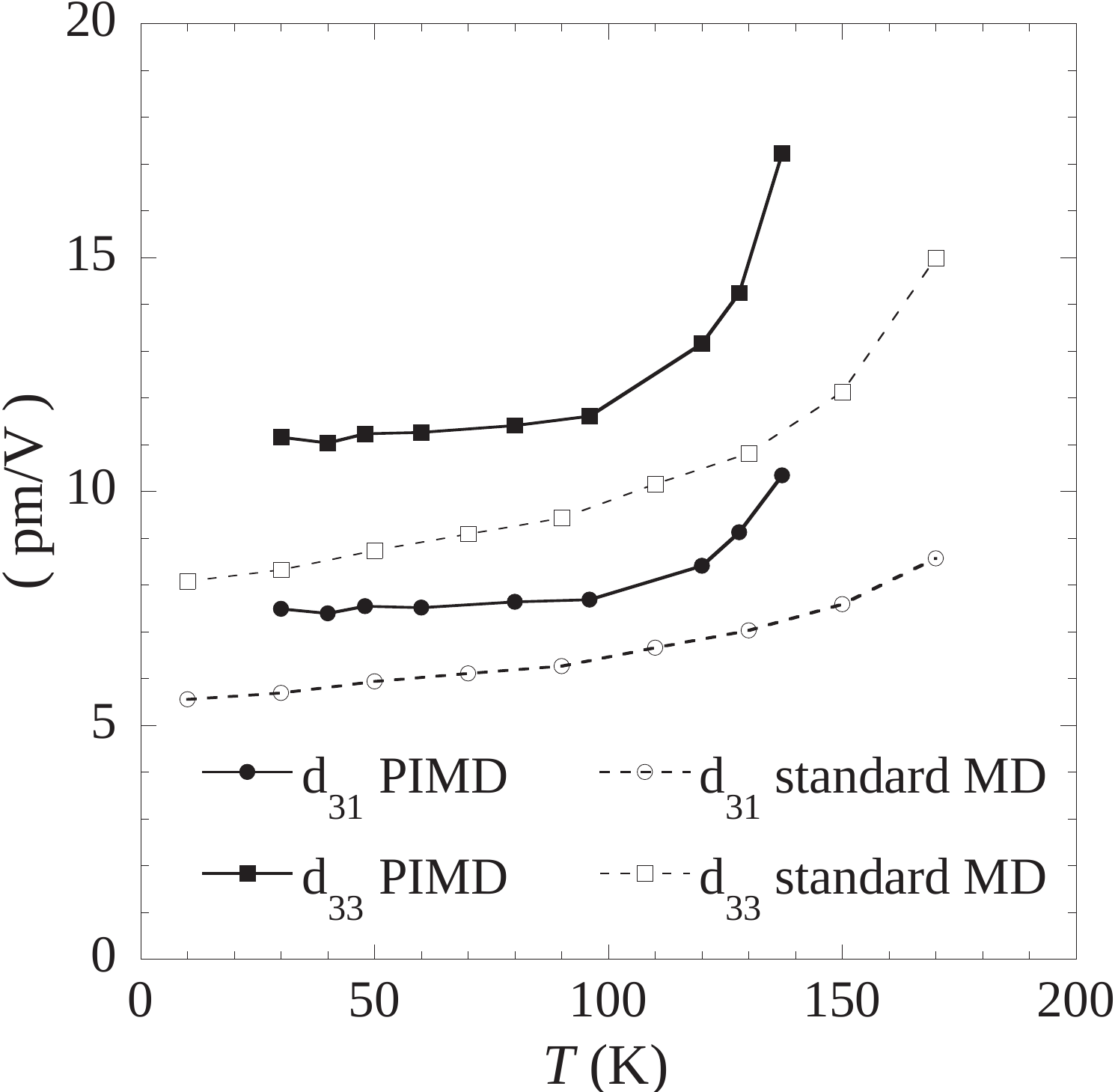}}}
    \par}
     \caption{{\small Temperature evolution of the $d_{31}$ and $d_{33}$ piezoelectric coefficients, computed using standard MD and by including the quantum effects through PIMD.}}
    \label{piezo3133}
\end{figure}

\begin{figure}[htbp]
    {\par\centering
     {\scalebox{0.5}{\includegraphics{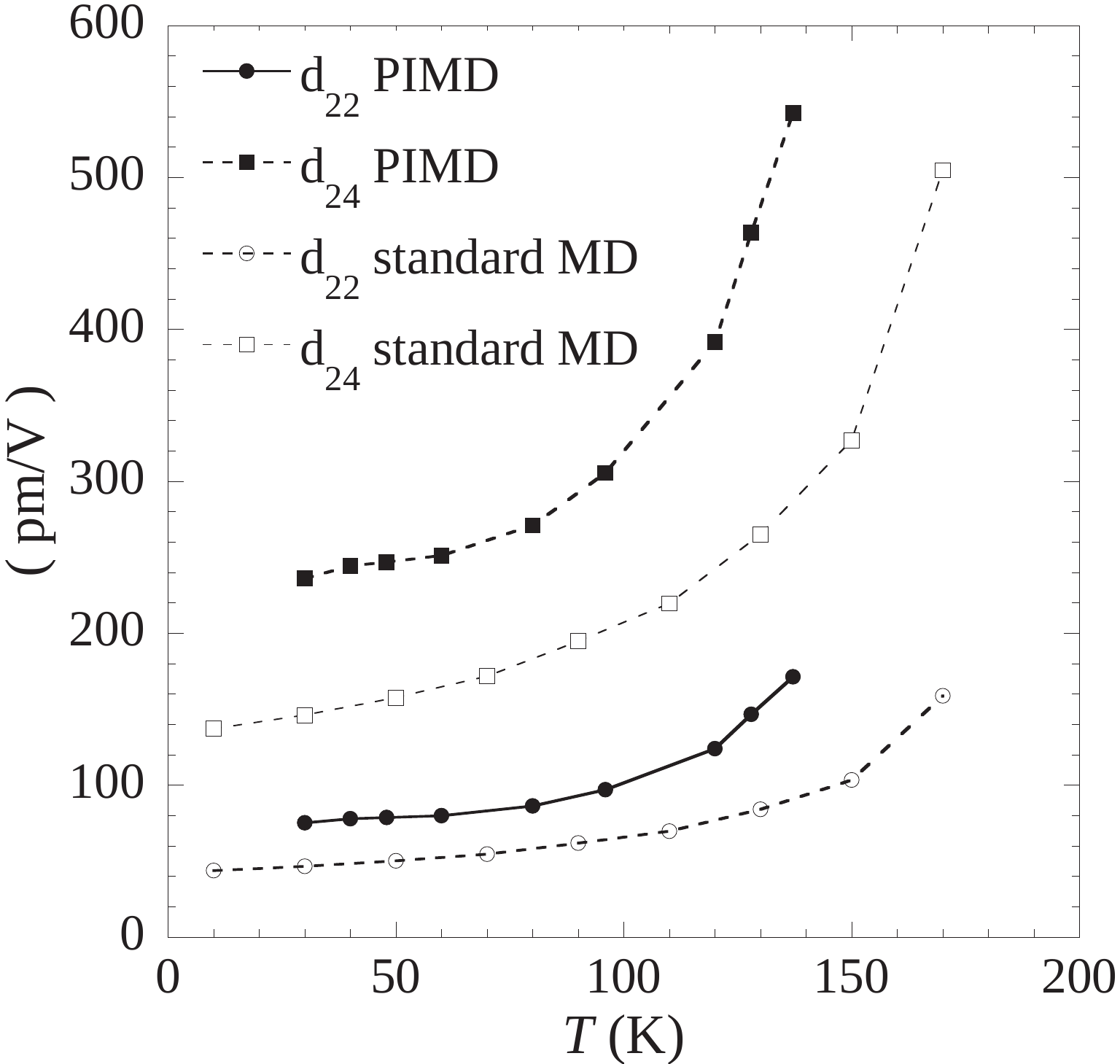}}}
    \par}
     \caption{{\small  Temperature evolution of the $d_{22}$ and $d_{24}$ piezoelectric coefficients, computed using standard MD and by including the quantum effects through PIMD.}}
    \label{piezo2224}
\end{figure}

Figures~\ref{piezo3133} and ~\ref{piezo2224} display the temperature evolution of the longitudinal, $d_{33}$, tranverse, $d_{31}$, and  shear, $d_{22}$ and $d_{24}$, piezoelectric constants in the R phase of BTO, computed in the classical and quantum-mechanical descriptions. The same types of effects are observed as for the dielectric constants. 

We note that, independently from the inclusion of quantum fluctuations, $d_{24}>d_{22}>d_{33}>d_{31}$ and the shear piezoelectric constants are one order of magnitude larger than the longitudinal and transverse constants. The values obtained at $T=30$~K are in good agreement with the DFPT calculations of Wu, Vanderbilt and Hamann\cite{wu} (performed at $T=0$~K). Note that the comparison with $T=0$~K results, reported in Tab.~\ref{TAB1}, is relevant since the properties slowly evolve with temperature below 100~K.

The components of the piezoelectric tensor have not been experimentally determined in the R phase, to the best of our knowledge. Nevertheless, the longitudinal constant along $[001]$, $d^*_{33}$, was derived from the slope of the strain versus electric field curves measured at $T=173$~K\cite{park} ($\approx 10$~K below the experimental R--O transition temperature). $d^*_{33}$ (see Appendix) calculated at $T=137$~K ($\approx 20$~K below the R--O transition temperature) is rather close to this experimental value (see Tab.~\ref{TAB1}).

\begin{figure}[htbp]
    {\par\centering
     {\scalebox{0.5}{\includegraphics{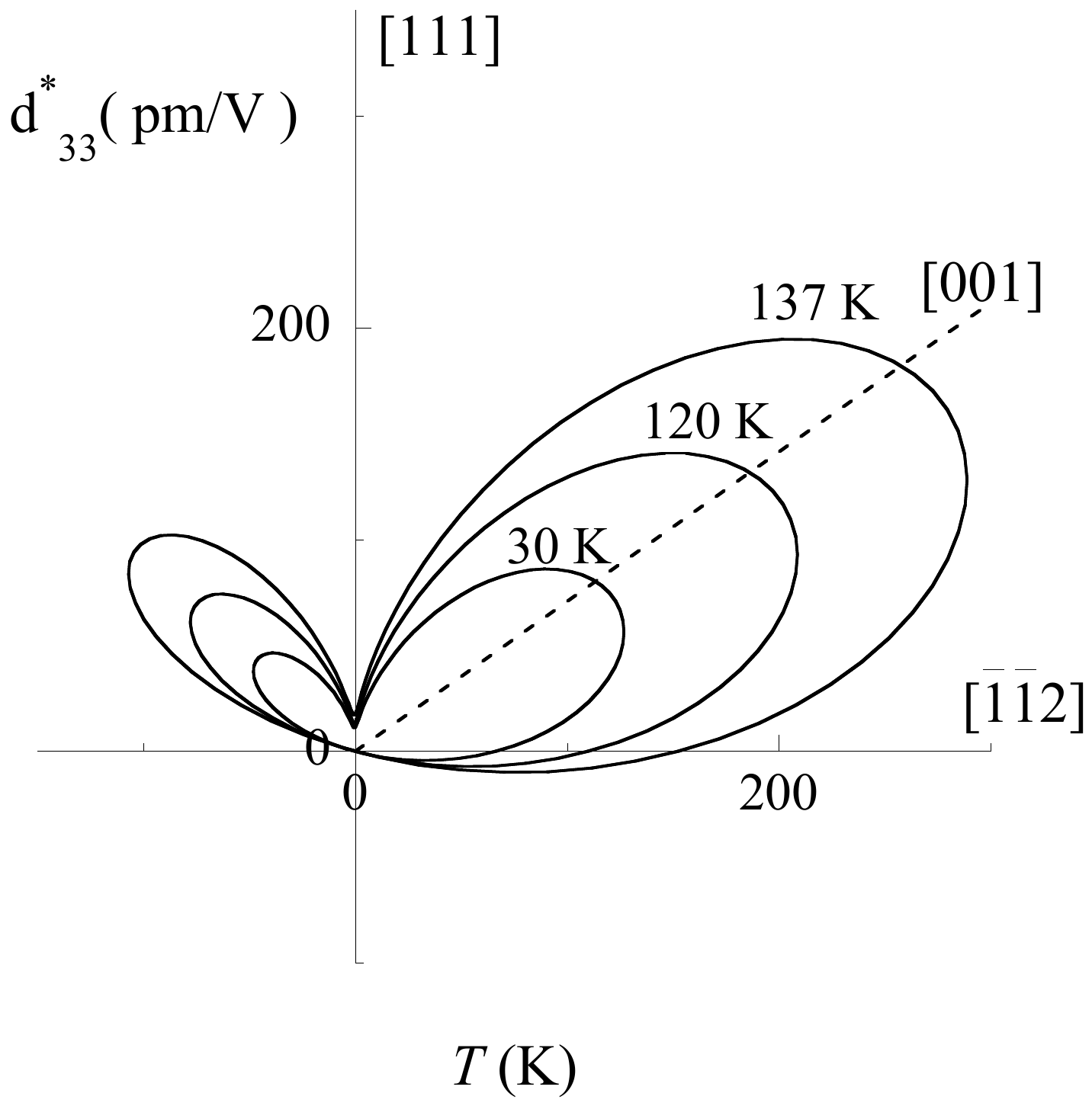}}}
    \par}
     \caption{{\small Orientational dependence of the $d_{33}^*$ piezoelectric coefficient  in the $(\bar{1}10)$ plane.}}
    \label{piezo001}
\end{figure}

It is important to mention that the longitudinal coefficient $d^*_{33}$, along $[001]$, is higher than $d_{33}$, along [111]. This strong anisotropy is due to the contribution of the super large shear piezoelectric constants, $d_{22}$ and $d_{24}$. Hence, it is interesting to calculate the orientation dependence of $d^*_{33}$ in order to identify the direction along which its value is the more enhanced. Figure ~\ref{piezo001} shows the orientation dependence of $d^*_{33}$ in the $(\bar{1}10)$ plane along which the maximum value is obtained. It corresponds to a direction close to $[001]$ and this behavior is clearly temperature-independent.

Hence, in the R phase of BTO the direction of enhanced piezoelectricity is different from the direction of the polarization. This feature was previously studied in giant-piezoelectric single crystals such as PMN-PT\cite{Damj03,Zhan03} or PZN-PT\cite{Damm03}. Such piezoelectric properties in $[001]$ oriented crystal has been attributed to the very large value of the $d_{24}$ shear coefficient. This property apparently specific to morphotropic compounds is thus observed in a simple perovskite like BTO. Indeed, in the case of BTO the $d_{24}/d_{33}$ ratio is around 21 and close to the ratio of 22 found for PMN-PT in its R phase.

\begin{table}[htdp]
\caption{Piezoelectic constants (pC/N) of barium titanate obtained by calculations.}
\begin{center}
\begin{tabular}{ccccccc}
  \hline
  \hline
                   & $T$(K)   & $d_{31}$   &$d_{33}$&$d_{22}$& $d_{24}$ & $d^*_{33}[001]$  \\
  \hline
present work       & $30$     & $ 7.5$     & $11  $ & $ 75 $ & $ 236  $ & $  137  $  \\
DFPT\cite{wu}      & $ 0$     & $ 6.8$     & $15  $ & $ 70 $ & $ 243  $ & $  137  $  \\
exp.\cite{park}    & $173$    &            &        &        &          & $  275  $  \\
present work       & $137$    &            &        &        &          & $  315\pm20  $  \\
  \hline
  \hline
\end{tabular}
\end{center}
\label{TAB1}
\end{table}

\section{Discussion}

It is useful to compare qualitatively the case of BTO to other ferroelectric systems. Thus we now discuss the importance of quantum fluctuations and their interplay with anharmonicities in such systems. We limit the discussion to standard FE systems, that do not exhibit any other order parameter than polarization, and to their lowest-temperature phase. Let us denote by $E^{zp}$ the zero-point energy associated to atomic motions, and by $\Delta V$ the typical (free) energy barrier to overcome to reverse the polarization (as would be given for instance by a phenomenological treatment within Landau theory). Different prototypical cases can be described.

(i) FE crystals with very deep double-well free energy landscape exhibit large barrier height ($\Delta V$) compared to $E^{zp}$ and thus a Curie temperature $T_c$ relatively high. The typical example of such systems is PbTiO$_3$. Upon cooling, for $T << T_c$, the crystal eventually behaves as an {\it harmonic system}. Therefore, its dielectric permittivity is expected to be rather flat at low temperature, and not sensitive to quantum effects. 

(ii) At the opposite, if $E^{zp} > \Delta V$ or if $\Delta V$ and $E^{zp}$ have similar values, the ground state is expected to extend over the different energy minima. In such cases, the ferroelectric transition might be suppressed by quantum zero-point fluctuations and the system remains paraelectric for all temperatures, despite the existence of a FE instability. This is typically the case of the quantum paraelectric crystals, such as KTaO$_3$.\cite{kto} In such systems, the dielectric permittivity increases upon cooling but, since the phase transition does not take place, it eventually saturates to a large value down to zero~K.

(iii) An intermediate situation corresponds to a zero-point energy $E^{zp} < \Delta V$, but large enough anyway so that the ground state extends up to the anharmonic region of the potential (while staying confined in a single energy minimum). In such a case, all the physical quantities exhibit therefore anharmonic behavior down to $T=0$~K. The dielectric permittivity, in particular, saturates at low temperature, as a property of the ground state. In the present work we have demonstrated that BTO belongs to this last case, {\it and is thus an anharmonic system down to zero Kelvin}. It is important to point out that in the case of the classical treatment, harmonic behavior is observed at sufficiently low $T$, leading here again to a plateau in the low-temperature evolution of the dielectric permittivity. However, the difference of origin between the quantum and classical plateaus is reflected by their different values (the quantum plateau is higher than the classical one).

\section{Conclusion}
In this work, the PIMD simulations have been performed to study low-temperature (R phase) dielectric and piezoelectric properties of BTO: spontaneous polarization, dielectric susceptibility and piezoelectric constants. We have shown that these three properties are different whether they are treated classically or quantum-mechanically. More precisely, significant enhancement of the dielectric tensor components and of the piezoelectric constants are observed as a consequence of the inclusion of quantum fluctuations. Such enhancement is attributed to the anharmonic contributions to the ground state, in which the system saturates at low temperature. 

By contrast, without including the quantum effects, the system eventually becomes harmonic at low temperature. In BTO, the polarization quantum-fluctuations in the ground state therefore extend over a region in which the potential energy surface is strongly anharmonic. This is corroborated by the significant difference between the low-temperature spontaneous polarizations computed classically and quantum-mechanically.

BTO is a strongly anharmonic system down to zero Kelvin, the anharmonicity being retained at low temperature by the quantum zero-point effects. The anharmonicity should thus be accounted for to achieve realistic predictions in this system, even in its low-temperature rhombohedral phase.

\appendix
\section{Piezoelectric-coefficient calculations}\label{PCC}
We choose the electric field {\bf E} and the stresses {\bf $\sigma$} as independent variables. The electromechanical properties are therefore described by the dielectric constants at constant stress $\epsilon_{ij}^\sigma$, and the piezoelectric constants $d_{i \alpha}$. Since there is no ambiguity, the superscripts $\sigma$ will be omitted for simplicity in the following.

The dielectric and piezoelectric matrix,  $\epsilon_{ij}$ and $d_{i \alpha}$,  of the rhombohedral single domain state using the [$\overline{1}$10], [$\overline{1}$$\overline{1}$2] and [111] axis and according to the $R3m$ symmetry is as below  

\[
\left[ {\begin{array}{ccc}
 \epsilon_{11} & 0 & 0  \\
 0 & \epsilon_{11} & 0  \\
 0 & 0 & \epsilon_{33} \\
 \end{array} } \right]
\]

\[
\left[ {\begin{array}{cccccc}
 0 & 0 & 0 & 0 & d_{24} & -2d_{22} \\
 -d_{22} & d_{22} & 0 & d_{24} & 0 & 0 \\
 d_{31} & d_{31} & d_{33} & 0 & 0 & 0 \\
 \end{array} } \right]
\]

The coefficients $\epsilon_{ij}^*$ and $d_{i \alpha}^*$ obtained by axis rotation, i. e. in the referential $[1 0 0]$, $[0 1 0]$ and $[0 0 1]$, are as below

\[
\left[ {\begin{array}{ccc}
 \epsilon_{11}^* & \epsilon_{31}^* & \epsilon_{31}^*  \\
 \epsilon_{31}^* & \epsilon_{11}^* & \epsilon_{31}^*  \\
 \epsilon_{31}^* & \epsilon_{31}^* & \epsilon_{11}^* \\
 \end{array} } \right]
\]

\[
\left[ {\begin{array}{cccccc}
 d_{33}^* & d_{31}^* & d_{31}^* & d_{36}^* & d_{34}^* & d_{34}^* \\
 d_{31}^* & d_{33}^* & d_{31}^* & d_{34} & d_{36}^* & d_{34}^* \\
 d_{31}^* & d_{31}^* & d_{33}^* & d_{34}^* & d_{34}^* & d_{36}^* \\
 \end{array} } \right]
\]

where 
\begin{equation} \label{EQA1}
 \epsilon_{31}^* = \frac{1}{3}\left(\epsilon_{33} - \epsilon_{11}\right)
\end{equation} 
\begin{equation} \label{EQA2}
 \epsilon_{33}^* = \frac{1}{3}\left(\epsilon_{33} +2 \epsilon_{11}\right)
\end{equation} 
\begin{equation} \label{EQA3}
 d_{31}^* = \frac{1}{3\sqrt{3}}\left(-\sqrt{2}d_{22} - d_{24} +2 d_{31} + d_{33}\right)
\end{equation} 
\begin{equation} \label{EQA4}
 d_{33}^* = \frac{1}{3\sqrt{3}}\left(2\sqrt{2}d_{22} + 2 d_{24} +2 d_{31} + d_{33}\right)
\end{equation}
\begin{equation} \label{EQA5}
 d_{34}^* = \frac{1}{3\sqrt{3}}\left(-2\sqrt{2}d_{22} + d_{24} -2 d_{31} + 2d_{33}\right)
\end{equation}
\begin{equation} \label{EQA6}
 d_{36}^* = \frac{1}{3\sqrt{3}}\left(4\sqrt{2}d_{22} -2 d_{24} -2 d_{31} + 2d_{33}\right)
\end{equation} 

We consider, at a given temperature, the rhombohedral phase  with a spontaneous polarization, {\bf P}$^0$ where $P_1^0=P_2^0=P_3^0$ and spontaneous homogenous strains, $\eta_\alpha^0$ where $\eta_1^0=\eta_2^0=\eta_3^0$ and $\eta_4^0=\eta_5^0=\eta_6^0$. When an electric field {\bf E} is applied along the [001] direction, the polarization and homogenous strain variations are given by 

\begin{equation} \label{EQA7}
 \Delta P_i=\epsilon_{3i}^*~E
\end{equation}
\begin{equation} \label{EQA8}
 \Delta \eta_{\alpha}=d_{3\alpha}^*~E
\end{equation}

From equilibrium trajectories for different values of the electric field, it is easy to derive the six dielectric and piezoelectric constants, $\epsilon_{ij}$ and $d_{i \alpha}$, by using equations (\ref{EQA1}-\ref{EQA8}).

\newpage

\end{document}